\documentclass[conference]{IEEEtran}
\usepackage{amsmath,amssymb}
\usepackage{graphicx}
\usepackage{booktabs}
\usepackage{array}
\usepackage{xcolor}
\usepackage{cite}
\usepackage[hidelinks,hypertexnames=false]{hyperref}
\usepackage[utf8]{inputenc}
\usepackage{algorithm}
\usepackage{algpseudocode}
\usepackage{microtype}
\usepackage{pgfplots}
\usepackage{newtxtext}
\pgfplotsset{compat=1.17}
\definecolor{fgmain}{HTML}{1F4E79}
\definecolor{fgmid}{HTML}{8FB4D9}

\begin{document}
\bstctlcite{IEEEexample:BSTcontrol}

\title{FPGA-GEN: Target-Aware RTL Generation via HLS}

\author{\IEEEauthorblockN{Luke Ztz Hu\textsuperscript{1}, Zhihan Xiao\textsuperscript{1}, Zhe Zhao\textsuperscript{1}, John Imoleayo Adebisi\textsuperscript{1}, Han Gao\textsuperscript{1}, Linxiao Yan\textsuperscript{1}, Songping Mai\textsuperscript{1}}
\IEEEauthorblockA{\textsuperscript{1}Shenzhen International Graduate School, Tsinghua University\\
Corresponding author: Songping Mai (\href{mailto:mai.songping@sz.tsinghua.edu.cn}{mai.songping@sz.tsinghua.edu.cn})}}

\maketitle

\begin{abstract}
Successful C/C++ compilation does not guarantee correct or efficient FPGA
hardware after high-level synthesis (HLS). FPGA-GEN is an agentic HLS-C
generation workflow that encodes specification, readiness, selection, repair,
and admission rules in explicit task contracts and bounded tool-feedback loops
for candidate-level decisions. On HLS-Eval, five shared-context outputs per
task yield 407/425 (95.8\%) public-testbench passes. In a cumulative conversion
campaign, FPGA-GEN admits 104/107 (97.2\%) HLS-LeVeri designs under held-out
co-simulation, Vivado synthesis/fit, and same-RTL evidence. In focused
TensorCore screening, selected candidate 3 reports 73.7\% fewer CSynth cycles
and 14.6\% fewer LUTs than generated/composed $C_0$, with 26.7\% more DSPs,
and passes development CoSim.
\end{abstract}

\section{Introduction}
\label{sec:introduction}
High-level synthesis (HLS) tools help convert HLS aware C/C++(HLSC) into
register-transfer level (RTL) designs for field-programmable gate
arrays (FPGAs) or application-specific integrated circuits
(ASICs)~\cite{5737854}. However, the generated RTL does not guarantee reference behavior or satisfy resource and performance constraints. Based on this, a useful FPGA hardware requires both functional validation and hardware-aware optimization: for example, a functionally correct implementation may exceed the resources
available on the target device or fail to meet application performance specifications~\cite{10.1145/3494534}.

This emergence of large language models (LLMs) provide a iterative solution to this problem. LLMs aided by repair policies can produce promising rtl code~\cite{liu2023verilogevalevaluatinglargelanguage,10.1145/3718088}. Workflows built upon LLMs have shown strong results on the HLS workflow generation.
In particular, C2HLSC refactors software C into HLS-compatible
code~\cite{10.1145/3734524}, while HLSRewriter combines guided
refactoring with hardware-oriented
optimization~\cite{10.1145/3749986}. HLSPilot integrates profiling,
C/C++-to-HLS translation, and pragma design-space
exploration~\cite{10.1145/3676536.3676781}. More recent agentic
workflows investigate iterative generation, tool-assisted refinement,
and validation in FPGA
environments~\cite{abikaram2026benchmarkingagentichlsdesign,
wang2026fpgagentllmassistedframeworkautonomous}.
These advances motivate workflows that connect generation and repair
to explicit application requirements and target-specific hardware
evidence.

A key issue still remains, how to generate performance acceptable functional equivalent HLS-C code. A candidate may synthesize
successfully while violating its behavioral contract or resource
budget. Conversely, reducing latency may increase resource use.
Repair iterations may further complicate evaluation because a shift in code can change both
behavior and hardware cost, making earlier measurements insufficient
to characterize the repaired implementation. Candidate selection and
acceptance must therefore identify the requirements, artifacts, and
evaluation conditions supporting each decision.

We present \textsc{FPGA-GEN}, a specification-guided workflow for
generating and repairing HLS-compatible C/C++ (HLS-C), with
target-aware candidate selection under a fixed search budget.
The workflow records task requirements, uses tool feedback to guide
repair, and distinguishes functional validation from device
admission. Its admission path associates functional and
device-synthesis evidence with matching RTL identities. A separate
bounded-selection path compares candidates using synthesis-derived
hardware cost before development validation and repair.
The contributions are as follows:
\begin{enumerate}
    \item \textbf{Multi-level task decomposition.}
    FPGA-GEN structures FPGA development at workflow and function levels.
    For reachable source helpers, bottom-up construction records unit-specific
    specifications, readiness checks, input-bound provenance and local
    verification status before top-level composition.

    \item \textbf{Bounded hardware-aware selection and repair.}
    In the focused study, FPGA-GEN screens a conservative reference and up to
    three model alternatives, ranks eligible candidates by a pre-repair
    latency--resource proxy, and validates and repairs the selected design
    within a fixed budget. A still-invalid model-selected design triggers
    reconstruction and a fresh bounded validation/repair loop for the
    conservative reference.

    \item \textbf{Artifact-linked output qualification.}
    FPGA-GEN distinguishes generated candidates, validated outputs, and
    device-admitted implementations. Its admission path requires held-out
    functional/RTL agreement and Vivado synthesis with resource fit on the
    same RTL. Missing or mismatched evidence prevents admission, and
    cost-selection results are reported separately.
\end{enumerate}

\begin{figure*}[!t]
    \centering
    \includegraphics[width=0.9\textwidth]{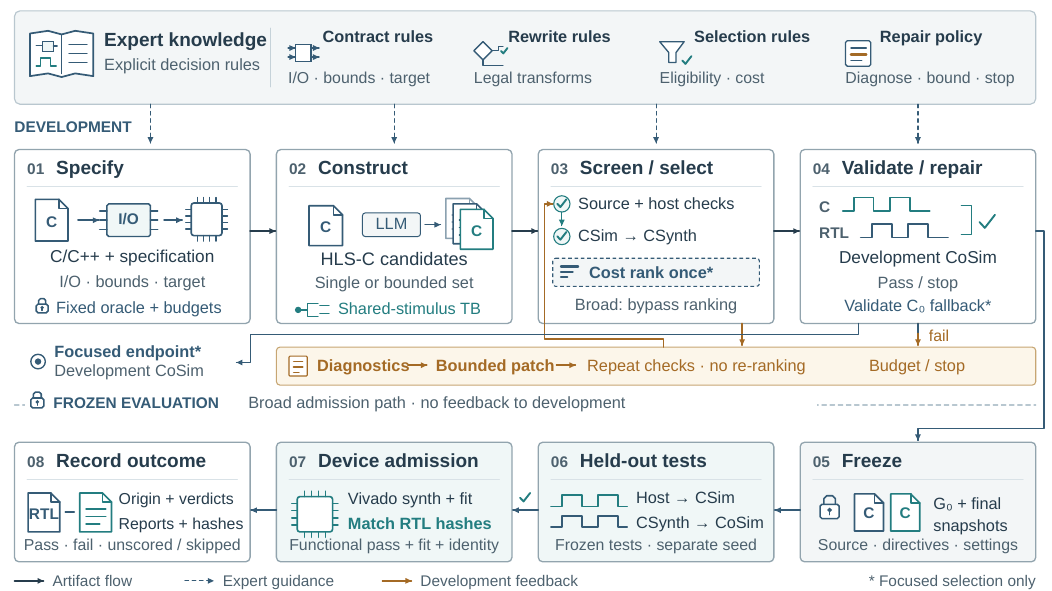}
    \caption{FPGA-GEN's rule-based workflow; inputs and testbenches follow the
    study protocol.}
    \label{fig:fpga-gen-workflow}
\end{figure*}

\section{Related Work}
\label{sec:related_work}

\subsection{C-to-HLS-C Translation and Generation}

Translating software C/C++ into synthesizable HLS-C removes unsupported
constructs while preserving behavior. HeteroRefactor uses dynamic invariants
~\cite{10.1145/3377811.3380340}; HeteroGen combines transformation, test
generation and repair~\cite{10.1145/3503222.3507748}. LLM systems add
in-context learning~\cite{10.1145/3676536.3676781}, hierarchical tool feedback
~\cite{10.1145/3734524}, repair templates~\cite{10.1145/3749986} and AST-guided
fine-tuning~\cite{khan2025sagehlssyntaxawareastguidedllm}. Zhao et
al.~\cite{zhao2026evidencedrivenllmagentctosynthesizablec} couple HLS-C
conversion to four-stage verification. HLStrans provides paired C and HLS
programs with testbenches and synthesis annotations for
C-to-HLS translation~\cite{zou2025hlstransdatasetctohlshardware}.

\subsection{Testbench Generation and Functional
Verification}

Synthesis does not demonstrate functional agreement. AutoBench generates HDL
testbenches, while CorrectBench studies their functional self-correction
~\cite{10.1145/3670474.3685956,Qiu2024CorrectBenchAT}. HLS-Eval
~\cite{11106033,abikaram2026benchmarkingagentichlsdesign} and Bench4HLS
~\cite{khan2026bench4hlsendtoendevaluationllms} span simulation and HLS
synthesis. Xiao et
al.~\cite{xiao2026shiftlefthighlevelsynthesisverification}
compare the C reference with its HLS-C version before co-simulation.
C/RTL co-simulation exercises the RTL, but finite tests establish
agreement only over the declared inputs~\cite{10.1145/3626202.3637563}.

\subsection{Tool-Guided Repair and Agentic
HLS Workflows}

Tool-guided agents use diagnostics for translation, repair and optimization
~\cite{luo2026hlsmithexpertguidedagenticframework,mastora2026executablebenchmarkllmbasedhls}.
HLSDebugger~\cite{wang2025hlsdebuggeridentificationcorrectionlogic},
ChatHLS~\cite{li2026chathlssystematicdesignautomation} and
AgRefactor~\cite{10.1145/3748173.3779566} add
specialized diagnosis, optimization and repair reasoning. MementoHLS combines
evidence-gated rules with per-candidate trajectory and outcome tracking
~\cite{zhu2026ruleguidedtrajectoryawarerepairllmbased}. Table~\ref{tab:closest-work}
compares the closest workflows by the capabilities their sources report.
Different inputs and protocols make this descriptive, not a ranking. Zhao et
al. join conversion, repair and verification under evidence isolation.
ContractHIL-HLS~\cite{zhang2026contracthilhlscontractalignedmultiagentworkflow}
feeds HLS, Vivado and board evidence into structured contracts.
FPGAgent~\cite{wang2026fpgagentllmassistedframeworkautonomous} combines
evolutionary validation, host-code generation and Alveo U200 execution.

\begin{table}[!b]
\centering
\caption{HLS Workflow Capabilities}
\label{tab:closest-work}
\footnotesize
\setlength{\tabcolsep}{2pt}
\begin{tabular}{@{}
>{\raggedright\arraybackslash}p{0.21\columnwidth}
>{\raggedright\arraybackslash}p{0.21\columnwidth}
>{\raggedright\arraybackslash}p{0.52\columnwidth}@{}}
\toprule
Workflow & Signal & Mechanism and endpoint \\
\midrule
Zhao et al.~\cite{zhao2026evidencedrivenllmagentctosynthesizablec}
& Software C/C++
& Four-stage conversion checks; closes at C/RTL CoSim. \\

MementoHLS\newline\cite{zhu2026ruleguidedtrajectoryawarerepairllmbased}
& Candidate-bound tool evidence
& Four checks bind public-testbench and synthesis outcomes to one candidate. \\

ContractHIL-HLS~\cite{zhang2026contracthilhlscontractalignedmultiagentworkflow}
& Natural-language contract
& HLS/Vivado feedback; HLS-Eval plus a separate board case. \\

FPGAgent~\cite{wang2026fpgagentllmassistedframeworkautonomous}
& Natural language
& Host/kernel generation and repair; FPGA board execution. \\

\textsc{FPGA-GEN}
& Task contract; reference C
& Held-out CoSim and same-RTL device fit; separate CSynth-cost selection. \\
\bottomrule
\end{tabular}
\end{table}

These workflows place evidence at different boundaries. MementoHLS binds four
public-testbench and synthesis checks to one candidate, while Zhao et al.
close conversion through C/RTL CoSim. ContractHIL-HLS reports its public
benchmark separately from a board case, and FPGAgent executes generated host
and kernel code on a board. FPGA-GEN instead separates CSynth cost screening
from held-out CoSim and same-RTL device-fit admission. Source generation,
tested RTL agreement, device feasibility and board execution are therefore
complementary endpoints, not interchangeable success measures.

\subsection{HLS Optimization and Design-Space Exploration}

Traditional HLS optimization explores code transformations and synthesis
directives to balance performance and resource use. AutoAccel combines an
accelerator template with analytical models to guide code
transformation~\cite{8465940}, while AutoDSE uses bottleneck analysis to
prioritize optimization parameters~\cite{10.1145/3494534}. More recently, SoberDSE
selects exploration algorithms using benchmark
characteristics~\cite{10.1145/3770743.3804049}, while HLS-Seek incorporates hardware quality
into model training through a comparative reward model and selective
synthesis feedback~\cite{zou2026hlsseekqorawarecodegeneration}.

\section{Methodology}
\label{sec:method}

\textsc{FPGA-GEN} maps task contracts and references to construction,
validation, repair and evaluation (Fig.~\ref{fig:fpga-gen-workflow}). The LLM
generates or repairs HLS-C, workflow code composes it, and Vitis HLS emits RTL.
G0 is the archived broad initial source. Focused $d^0$ is the selected eligible
candidate or C0 fallback. Helpers are callees below the synthesis top. The
returned design is the final frozen HLS-C. Generation, validation and admission
remain distinct.

\subsection{Task Contract, Readiness and Hierarchical Construction}
\label{sec:method_contract}

A task fixes reference C/C++ $g_i$; top function, interface, bounds and state;
valid inputs and comparisons; FPGA/clock; and budgets. A missing oracle,
required dependency/interface or test bound blocks the task. Finite tests
establish only tested behavior.

Reachable helpers are processed leaves-first, recording local specifications,
readiness and bound provenance before composition. Matching oracles and bounds
enable local validation and repair; otherwise helpers remain unverified or
failed. Composition promotes neither to pass, and the top-level ladder against
$g_i$ decides the verdict. With no helpers, construction proceeds directly.

The conservative baseline $d_i^{\mathrm{C0}}$ retains the top body and
signature with configured pragmas but shares the processed, possibly generated
helpers used by alternatives. Thus C0 need not be wholly original: $g_i$
defines correctness and screened C0 the QoR reference.

Generation uses only protocol-exposed source and HLS context. The reference
and candidate share test vectors~\cite{xiao2026shiftlefthighlevelsynthesisverification}.
Host tests compare observables (runnable vector-add example). CSim executes
C/C++ and CoSim exercises synthesized RTL~\cite{amd2024hls}. These establish
tested agreement, not formal equivalence.

\subsection{Bounded Hardware-Aware Candidate Selection}
\label{sec:method_generation}

Broad mode generates one candidate and archives source as G0 before validation
or repair; helper construction separates model-call and candidate counts.
Focused mode requests up to three alternatives and constructs C0 with the same
helpers. Source/interface checks, host comparison, CSim and CSynth screen the
finite pool; passing all stages defines eligibility.

For a candidate that passes screening and has usable reports, the
selection objective is
\begin{equation}
    J_i(d)=
    \frac{T_i(d)}{T_i(b_i)}
    \frac{1}{|R_i|}
    \sum_{r\in R_i}
    \frac{U_{ir}(d)}{U_{ir}(b_i)},
    \label{eq:method_dse}
\end{equation}
where $T_i(d)$ is worst-case CSynth cycles times estimated clock period---a
synthesis delay proxy, not CSim runtime, measured execution or routed timing---
and $U_{ir}(d)$ is reported resource use. The reference $b_i$ is C0 when its
screening and QoR reports are valid, otherwise the first screened point with
valid QoR. $R_i$ contains LUT, FF, DSP and BRAM classes with known, nonzero
reference use. Their normalized mean is a resource proxy, not physical area;
the cost is finite only with positive finite reference delay, nonempty $R_i$
and known candidate/reference counts.

Selection minimizes $J_i$ under common part/clock/tool settings; unavailable
cost is $+\infty$ only for selection. Ties prefer model candidates then
generation order, without establishing gain. C0 proceeds when no alternative
or eligible point exists; no source stops the run. Finite best-of-$N$ search
does not establish a global optimum.

\subsection{Development Validation, Repair and Conservative Recovery}
\label{sec:method_repair}

Only the selected focused candidate or single broad candidate enters the full
development ladder. The earliest actionable diagnostic guides revision, and
changed source restarts validation. Broad evaluation allows three repairs;
focused evaluation allows three validations and two repairs. Pass, repeated
source, nonactionable feedback or exhaustion stops the loop. Timeouts, tool
failures and missing reports remain unscored, not mismatches.

Search ranks once before repair and does not recost changed source. An invalid
model winner triggers C0 reconstruction with the same helpers and a fresh
loop; its actual verdict is returned. Source, directives, target, settings and
origin are frozen (Algorithm~\ref{alg:method_workflow}). Focused evaluation
ends at development CoSim, before broad admission.

\begin{algorithm}[t]
\caption{FPGA-GEN bounded workflow}
\label{alg:method_workflow}
\begin{algorithmic}[1]
\Require Task $i$; mode $m$; endpoint $e$; declared budgets
\Ensure Source origin, stage verdicts, QoR and admission evidence
\State Check contract readiness; record a blocking outcome and stop if needed
\State Process reachable helpers leaves-first; retain local status and bounds
\If{$m=\mathrm{single}$}
    \State $d^0 \gets \Call{GenerateAndArchive}{i}$
\Else
    \State $D \gets \Call{GenerateUpToThree}{i}$
    \State $d^{\mathrm{C0}} \gets \Call{ConservativeBaseline}{i}$; $d^0\gets d^{\mathrm{C0}}$
    \If{$D\neq\varnothing$}
        \State $E \gets \Call{ScreenAndSynthesize}{\{d^{\mathrm{C0}}\}\cup D,i}$
        \State If $E\neq\varnothing$, $d^0 \gets \Call{Select}{E,J_i}$
    \EndIf
\EndIf
\State If no source exists, record the outcome and stop
\State $(d,v) \gets \Call{ValidateAndRepair}{d^0,i}$
\If{$m=\mathrm{search}$, $d^0$ is model-generated and $v\neq\mathrm{PASS}$}
    \State Rebuild C0 with the same helpers; $(d,v) \gets \Call{FreshC0Loop}{i}$
\EndIf
\State \Call{FreezeAndRecord}{d,v,$\mathrm{origin}(d)$, settings}
\If{$e=\mathrm{admission}$}
    \ForAll{distinct snapshots $q\in Q_i$}
        \State $(F,Z,h^H,h^V) \gets \Call{HeldOutThenDevice}{q,i}$
        \State \Call{RecordAdmission}{$q,A_i(q),F,Z,h^H,h^V$}
    \EndFor
\EndIf
\end{algorithmic}
\end{algorithm}

\subsection{Held-Out Evaluation and Artifact-Linked Admission}
\label{sec:method_admission}

For broad admission, $Q_i$ contains archived G0 and each distinct returned
snapshot. Frozen snapshots are checked separately; unchanged source may reuse
only identity-matched results. Vectors are fixed before generation and withheld
from development. They share the generator, oracle and directed prefix but use
a new random suffix, preventing feedback without claiming independent coverage.

Let $F_i(d)$ denote all required held-out functional/RTL passes and $Z_i(d)$
successful device synthesis with a valid fit report:
\begin{equation}
    A_i(d)=F_i(d)\land Z_i(d)\land
    \bigl[h_i^{H}(d)=h_i^{V}(d)\neq\varnothing\bigr],
    \label{eq:method_admission}
\end{equation}
where $h_i^H$ and $h_i^V$ identify the CoSim and synthesis RTL. Test counts
remain stage-specific. Missing or mismatched evidence prevents admission
without proving a counterexample. Device fit is Vivado synthesis within device
capacity, not placement, routability, timing closure or board execution; the
clock is a synthesis setting. Evaluation does not feed back.

\subsubsection{Endpoint Consistency Across Planned Outputs}
Public-profile passage and artifact admission are distinct row predicates. For
planned output $d_{ij}$, define
\begin{equation}
 P_{ij}^{\mathrm{pub}}=E_{ij}\land B_{ij}\land T_{ij}\land S_{ij},\qquad
 \mathrm{Pass}_{ij}^{\wedge}:=A_i(d_{ij}),
 \label{eq:method_consistency_endpoints}
\end{equation}
where $E$, $B$, $T$ and $S$ denote extractable source, public-testbench build,
public-testbench exit zero and HLS synthesis for the same source-bound output.
For a declared fixed set of $k$ planned outputs,
\begin{equation}
 \mathrm{All}_{i}^{(k)}(X)=\bigwedge_{j=1}^{k}X_{ij},\qquad
 \mathrm{Any}_{i}^{(k)}(X)=\bigvee_{j=1}^{k}X_{ij}.
 \label{eq:method_consistency_aggregates}
\end{equation}
$\mathrm{All}$ measures all-$k$ consistency and $\mathrm{Any}$ empirical
at-least-one coverage. For a declared planned population of $M$ tasks,
\begin{equation}
 R_G^{(k)}(X)=\frac{1}{M}\sum_{i=1}^{M}
 \mathbf{1}[G_i^{(k)}(X)=1],\quad G\in\{\mathrm{All},\mathrm{Any}\}.
 \label{eq:method_consistency_rates}
\end{equation}
Every planned slot remains in the denominator; unresolved required evidence
makes the rate unavailable. Broad $k=5$ indexes shared-context outputs in
generation order, not the focused C0-plus-three pool ($N\leq4$) ranked by
Eq.~\ref{eq:method_dse}; neither is an unbiased Pass@$k$ estimator. Initial and
repaired snapshots remain distinct. Missing admission evidence is unknown, and
\textsc{No\_Source} denotes no artifact, not a mismatch. Full-85
$X=\mathrm{Pass}^{\wedge}$ needs complete candidate-level admission rows.

\section{Experimental Results}
\label{sec:results}

\subsection{Experimental Setup}
Table~\ref{tab:studies} separates nonpooled profiles. HLS-Eval judges five
shared-context outputs per task on the public profile without repair or
admission. Focused mode screens C0 plus up to three alternatives and gives only
the selected or fallback return a development CoSim verdict. Conversion and
repair use held-out and admission endpoints.
Conversion draws
from seven sources, including 107 HLS-LeVeri designs and its test generator
~\cite{xiao2026shiftlefthighlevelsynthesisverification,1855741.1855756};
repair reuses frozen initial candidates. The 214-output, 49-task replay is
nested, not pooled. For the 85-task public profile, Codex helped recount passing
outputs and tasks from archived attempts using
Eqs.~\ref{eq:method_consistency_endpoints}--\ref{eq:method_consistency_rates}.
Separate records cover eight Rosetta tasks. The 70-design analysis and Rosetta
overlap on four designs (union 74), but unknown shared identities keep
denominators apart.

\begin{center}
\begin{minipage}{\linewidth}
\makeatletter\def\@captype{table}\makeatother
\centering
\caption{Studies and Endpoints}
\label{tab:studies}
\footnotesize
\setlength{\tabcolsep}{3pt}
\begin{tabular}{@{}>{\raggedright\arraybackslash}p{0.17\columnwidth}>{\raggedright\arraybackslash}p{0.25\columnwidth}>{\raggedright\arraybackslash}p{0.30\columnwidth}>{\raggedright\arraybackslash}p{0.18\columnwidth}@{}}
\toprule
Study & Tasks & Path & Endpoint \\
\midrule
Initial generation & 85 HLS-Eval~\cite{11106033}, five attempts & Generate $\to$ public TB; CSynth & Public TB \\
Repair, held-out & 70 of those designs; 8 Rosetta & Generate $\to$ repair $\to$ held-out & Functional; Rosetta admission \\
Conversion & 208 C tasks & Generate $\to$ repair $\to$ held-out CoSim $\to$ device fit & Cumulative admission \\
Focused selection & 32 incl. TensorCore & Generate $\to$ screen/CSynth $\to$ select/repair & Development CoSim \\
\bottomrule
\end{tabular}
\end{minipage}
\end{center}

Studies use Vitis HLS/Vivado 2024.2. HLS-Eval uses
\texttt{gpt-6.1-sol}/highest on U200
(\texttt{xcu200-fsgd2104-2-e}, 5\,ns); the original 208-task conversion used
author-reported \texttt{Opus 5.5}/medium, while cumulative-rerun model identity
is unrecorded. Targets/clocks are task-recorded. Admission requires held-out
C/RTL CoSim and same-RTL device fit, not routed timing or board results.

\subsection{Initial Generation and Public Consistency}
For 85 tasks with five attempts each, Table~\ref{tab:hlseval} reports
candidate- and task-level endpoints separately; \emph{Designs} counts tasks
with at least one passing attempt.
\begin{center}
\begin{minipage}{\linewidth}
\makeatletter\def\@captype{table}\makeatother
\centering
\caption{HLS-Eval Generation Outcomes}
\label{tab:hlseval}
\begin{tabular*}{\columnwidth}{@{\extracolsep{\fill}}lrrrr@{}}
\toprule
Stage & Candidates & (\%) & Designs & (\%) \\
\midrule
Code extraction  & 425 & 100.0 & 85 & 100.0 \\
Public-TB build  & 417 &  98.1 & 84 &  98.8 \\
Public testbench & 407 &  95.8 & 83 &  97.6 \\
HLS synthesis    & 425 & 100.0 & 85 & 100.0 \\
\bottomrule
\end{tabular*}
\end{minipage}
\end{center}

All 425 candidates extract and synthesize; 417 public testbenches build and 407
pass. Eight AES/DES candidates omit a helper, and ten fail NaN propagation,
Taylor-series sine or Needleman--Wunsch. These candidates are unrepaired, and
public passage is not C/RTL admission.

\subsubsection{Five-Output Endpoint Consistency}
Table~\ref{tab:consistency} applies Eqs.~\ref{eq:method_consistency_endpoints}--
\ref{eq:method_consistency_rates} to declared five-output cohorts and snapshots.
For the initial 85-task public profile, 18 failures are concentrated in five
tasks: 80 pass all five outputs, three are mixed and two pass none. Thus the
pooled 407/425 passage rate does not by itself express per-task consistency.

\begin{table}[!b]
\centering
\caption{Five-Output Consistency}
\label{tab:consistency}
\footnotesize
\begin{tabular*}{\columnwidth}{@{\extracolsep{\fill}}llrrr@{}}
\toprule
Cohort & Snapshot, $X$ & Slots & $\mathrm{Any}^{(5)}$ & $\mathrm{All}^{(5)}$ \\
\midrule
HLS-Eval (85) & Initial, $P^{\mathrm{pub}}$ & 407/425 & 83/85 & 80/85 \\
Rosetta (8) & Returned, $\mathrm{Pass}^{\wedge}$ & 40/40 & 8/8 & 8/8 \\
\bottomrule
\end{tabular*}
\end{table}

All 40 returned Rosetta outputs pass held-out CoSim and same-RTL admission, so
both task aggregates are 8/8. Sealed-finalizer lineage supports the joins;
compact records omit repeated RTL hashes, and the 40 raw pairs were not all
rehashed. Initial 35/40 is held-out only, not $\mathrm{Pass}^{\wedge}$.
Full-85 admission consistency needs its candidate ledger. Shared-context,
generation-order outputs are neither independent nor QoR-ranked samples.

\subsection{Repair and Held-Out Checking}
Repair without selection evaluates frozen candidates on judge set $J$ and
fresh-seed set $T$. Of 350 candidates, 325 are unchanged and 25 repaired. Each
set has 298 pass, 9 fail and 43 unscored; jointly, 296/9/45, so all 305 scored
candidates agree. Both sets bind the same returned source-tree hashes. Their
shared generator/reference makes this seed-repeatability, not independent
validation.

All nine failures are unrepaired (five AES, four NaN); all five
Needleman--Wunsch attempts are repaired and pass. Forty-five timeouts (27
PolyBench, 16 MachSuite, two C2HLSC-derived) bound per-set passage at
85.1--97.4\%. Each set runs up to 100 software, 100 CSim and 10 or 100 CoSim
tests. The nine failures pass this ladder but fail the public testbench;
ladder-only totals are 307/0/43. Reference tests do not subsume the benchmark.

The control has $J=(275,9,66)$ and $T=(291,16,43)$ for
(pass, fail, unscored). With 12.3\% unscored, the $T$-arm pass-rate difference
is bounded by $[-14.3,10.3]$ points. Thus the arms establish neither an
isolation advantage nor equivalence; these are functional passes, not
admissions.

Table~\ref{tab:repair-cases} records two repair cases and an
identity-preserved frozen replay.
\begin{table}[!b]
\centering
\caption{Repair and Replay Cases}
\label{tab:repair-cases}
\footnotesize
\begin{tabular}{@{}>{\raggedright\arraybackslash}p{0.20\columnwidth}>{\raggedright\arraybackslash}p{0.31\columnwidth}>{\raggedright\arraybackslash}p{0.39\columnwidth}@{}}
\toprule
Case & Initial observation / scope & Change / outcome \\
\midrule
Rosetta outer product & 5/5 initial slots fail held-out software checks &
5/5 returned sources change; cohort 35/40$\rightarrow$40/40 \\
Needleman--Wunsch, routed & Initial public TB fails; parse/build/synth pass &
Traceback ties: diagonal-first $\to$ left/up/diagonal; held-out: 99 pass, 1 skipped \\
Frozen replay & 214 checks on a selected 49-task cohort; no generation or repair &
159 pass; 1 mismatch; 44 timeout; 10 insufficient-vector \\
\bottomrule
\end{tabular}
\end{table}

Table~\ref{tab:suites} reports separate archived R2 terminal-functional
evidence from an asserted common 350-candidate start.
\begin{table}[!b]
\centering
\caption{Repair Cost and Coverage}
\label{tab:suites}
\footnotesize
\setlength{\tabcolsep}{2.5pt}
\begin{tabular*}{\columnwidth}{@{\extracolsep{\fill}}lrrrrr@{}}
\toprule
Policy & Pass/fail/unsc. & Repairs & Vitis & Vivado & \shortstack{Attempt\\time (s)} \\
\midrule
Plain  & 328/10/12 & 97 & 159 & 53 & 218{,}425.1 \\
Routed & 296/10/44 & 25 &  48 & 21 &  34{,}997.6 \\
\bottomrule
\end{tabular*}
\end{table}

In the current cohort, repaired sources are 23/25 pass with two unscored
(92--100\% bound); unchanged sources are 273/325 pass, nine fail and 43
unscored (84.0--97.2\%). These missingness bounds are not confidence intervals,
and selective triggering makes the groups non-causal. The joint verdict also
includes the public testbench and is not admission.

All five failed Rosetta outer-product slots change, although the records used
here do not represent exact patches. Held-out passage rises 35/40 to 40/40
with same-RTL admission; each repair records 100 software, 100 CSim and 10
CoSim tests. The control also reaches 40/40, supporting repair but not an
isolation advantage.

In frozen replay, native CSynth precedes RTL replay. All 214 rows carry hashes
with unchanged source, testbench and RTL and no generation or repair. The only
failure is \texttt{HE-machsuite-gemm\_blocked} sample 5. This 49-task cohort
supports identity-preserved replay, not benchmark-wide success or admission.

Archived R2 is not R6 joint $J/T$: both arms use an asserted 350-candidate
aggregate, \texttt{gpt-6.1-sol}/highest and at most three repairs, without
uniform served-identity verification or a row ledger. Routed repair uses 25
versus 97 calls and 84.0\% less summed completed-attempt time (34,997.6 versus
218,425.1~s), while 44 versus 12 outcomes are unscored. This is descriptive
work/coverage, not elapsed speedup or focused selection. Vitis/Vivado count
verification ladders. Of 63 selected pairs, 22 are jointly scored (21 both
pass, one plain-only). This is not a correctness, causal or significance
result.

\subsection{Conversion and Device Admission}
Table~\ref{tab:coverage} reports the cumulative best-per-task campaign:
\emph{Gen.} counts tasks producing source, and \emph{Adm.} counts
author-reported admissions under Eq.~\ref{eq:method_admission}. Rosetta is
incomplete, so no suite rate is reported.
\begin{center}
\begin{minipage}{\linewidth}
\makeatletter\def\@captype{table}\makeatother
\centering
\caption{Cumulative Device Admissions}
\label{tab:coverage}
\setlength{\tabcolsep}{5.5pt}
\begin{tabular*}{\linewidth}{@{\extracolsep{\fill}}lrrrr@{}}
\toprule
Cohort & Tasks & Gen. & Adm. & Adm. (\%) \\
\midrule
HLS-LeVeri~\cite{xiao2026shiftlefthighlevelsynthesisverification} & 107 & 104 & 104 & 97.2 \\
C2HLSC-derived~\cite{10.1145/3734524} & 9 & 9 & 9 & 100\\
Development (internal) & 32 & 32 & 30 & 93.8 \\
HLSRewriter-derived~\cite{10.1145/3749986} & 24 & 19 & 17 & 70.8 \\
MachSuite~\cite{6983050} & 19 & 18 & 16 & 84.2 \\
CHStone~\cite{4541637} & 12 & 10 & 9 & 75.0 \\
Rosetta (partial)~\cite{10.1145/3174243.3174255} & 5 & 5 & 5 & N/A \\
\midrule
Total & 208 & 197 & 190 & 91.3 \\
\bottomrule
\end{tabular*}
\end{minipage}
\end{center}
Across runs, including six HLS-LeVeri reruns, the author-reported aggregate
produces source for 197/208 tasks and admits 190 (91.3\% overall; 96.4\% of
generated). HLS-LeVeri reports 104/107 admitted and three without source.
These are not fixed-budget success estimates.

A separate follow-up revisited three LeVeri tasks without changing
Table~\ref{tab:coverage}'s totals. Two references block generation; the remaining candidate
passes 100 hidden CSim tests and CSynth, then RTL CoSim times out after 3 of 10
transactions. CSynth estimates 512 BRAMs against 280 available, and Vivado is
never launched, separating software passage, RTL completion and device
feasibility.

Of 18 non-admissions, 11 lack source (three HLS-LeVeri, five HLSRewriter, one
MachSuite, two CHStone) due to invalid oracles, dependencies or unresolved
tops. Seven generated tasks are not admitted (two internal, two HLSRewriter,
two MachSuite, one CHStone); neither status alone proves a counterexample.

\subsection{Focused Selection and TensorCore}
\begin{figure}[t]
\centering
\begin{tikzpicture}
\begin{axis}[
    xbar stacked, bar width=9pt, width=\columnwidth, height=2.5cm,
    xmin=0, xmax=32, xtick={0,8,16,24,32},
    xlabel={Tasks}, ytick=\empty, ymin=-0.6, ymax=0.6,
    axis line style={gray!60}, tick style={draw=none},
    tick label style={font=\footnotesize}, label style={font=\footnotesize},
    title={Selection-time outcome, 32 tasks}, title style={font=\footnotesize, yshift=9pt},
    legend style={at={(0.5,1.02)}, anchor=south, legend columns=4,
        draw=none, font=\footnotesize,
        /tikz/every even column/.append style={column sep=4pt}},
    legend image code/.code={\fill[#1] (0cm,-0.07cm) rectangle (0.22cm,0.11cm);},
    nodes near coords, point meta=explicit symbolic,
    every node near coord/.append style={font=\footnotesize, anchor=east, inner sep=1.5pt},
]
\addplot[fill=fgmain, draw=white, line width=0.5pt, text=white] coordinates {(21,0) [21]};
\addplot[fill=fgmid, draw=white, line width=0.5pt, text=black] coordinates {(7,0) [7]};
\addplot[fill=gray!45, draw=white, line width=0.5pt, text=black] coordinates {(3,0) [3]};
\addplot[fill=gray!18, draw=white, line width=0.5pt] coordinates {(1,0) [1]};
\legend{Lower cost, Equal cost, $C_0$ kept, Unresolved}
\end{axis}
\end{tikzpicture}
\caption{Focused study: 32-task selection outcomes before repair.}
\label{fig:focused}
\end{figure}
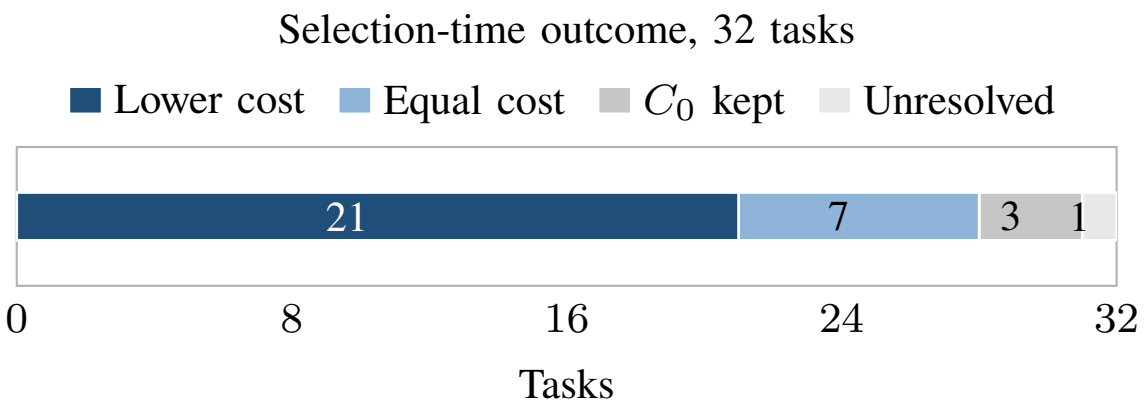

Focused selection ranks $C_0$ and up to three alternatives by
Eq.~\ref{eq:method_dse}. The checkpoint records 96 model candidates reaching
CSynth; only selected or fallback designs proceed to development CoSim. Of 32
tasks, 31 pass development CoSim; alternatives
are selected 28 times (21 lower cost, seven ties; Fig.~\ref{fig:focused}).
Three retain $C_0$ (\texttt{des\_\allowbreak permute},
\texttt{gemm\_\allowbreak 8x8}, \texttt{stencil2d}); \texttt{quantize} is
unresolved due to reference undefined behavior. Costs precede repair. For
\texttt{gemm\_\allowbreak 8x8}, an alternative cuts cycles 259 to 136 but has
cost 1.48, retaining $C_0$. A selector effect requires selected and $C_0$ arms
from the same pool/target/budget at one independent final endpoint; the
reported focused evidence does not supply that comparison.

Table~\ref{tab:tensorcore} compares selection-stage CSynth metrics for the
generated/composed shared-helper $C_0$ and selected candidate 3.
\begin{table}[!b]
\centering
\caption{TensorCore Cycles and Resources}
\label{tab:tensorcore}
\footnotesize
\begin{tabular*}{\columnwidth}{@{\extracolsep{\fill}}lrrr@{}}
\toprule
Metric & $C_0$ & Candidate 3 & Change (\%) \\
\midrule
CSynth cycles & 2,081 & 547   & $-73.7$ \\
LUTs          & 5,825 & 4,973 & $-14.6$ \\
FFs           & 6,232 & 6,244 & $+0.2$ \\
DSPs          & 75    & 95    & $+26.7$ \\
\bottomrule
\end{tabular*}
\end{table}

Candidate 3 (cost 0.274) passes development CoSim, trading lower CSynth
cycle/LUT counts for more DSPs and near-unchanged FFs. This profile ends at
development CoSim; referenced KV260 and retargeted runs lack matched receipts
in the manuscript evidence.

\subsection{Comparison with Published Workflows}
Table~\ref{tab:agent-comparison} places reported rates beside their task sets
and endpoints; \emph{all-pass} follows the criterion of
Zhao et al.~\cite{zhao2026evidencedrivenllmagentctosynthesizablec}. Agentic
HLS-Eval's gpt-oss-120b exceeds 90\% public-testbench pass@1 and reaches 100\%
at pass@10; the table shows the latter. Zhao et al. use CFull107; FPGA-GEN uses
HLS-LeVeri, so equal sizes do not imply matched tasks.
\begin{center}
\begin{minipage}{\linewidth}
\makeatletter\def\@captype{table}\makeatother
\centering
\caption{Benchmark Results by Workflow}
\label{tab:agent-comparison}
\footnotesize
\renewcommand{\arraystretch}{0.96}\setlength{\tabcolsep}{2.5pt}
\begin{tabular}{@{}>{\raggedright\arraybackslash}p{0.57\columnwidth}>{\raggedright\arraybackslash}p{0.25\columnwidth}>{\raggedleft\arraybackslash}p{0.10\columnwidth}@{}}
\toprule
\textbf{Workflow} & \textbf{Endpoint} & \textbf{Rate (\%)} \\
\midrule
\multicolumn{3}{@{}l@{}}{\textbf{HLS-Eval generation} (85 tasks)} \\
\hspace{0.5em}DeepSeek V3, zero-shot~\cite{11106033} & Public TB, pass@1 & 63.3 \\
\hspace{0.5em}Agentic HLS-Eval, gpt-oss-120b~\cite{abikaram2026benchmarkingagentichlsdesign} & Public TB, pass@10 & 100.0 \\
\hspace{0.5em}\textsc{FPGA-GEN}, initial & Pooled public TB & 95.8 \\
\multicolumn{3}{@{}l@{}}{\textbf{ChatHLS generation} (108 tasks; 20 samples/task)} \\
\hspace{0.5em}RAG~\cite{li2026chathlssystematicdesignautomation} & CoSim, pass@1 & 52.3 \\
\hspace{0.5em}HLSFixer~\cite{li2026chathlssystematicdesignautomation} & CoSim, pass@1 & 77.2 \\
\midrule
\multicolumn{3}{@{}l@{}}{\textbf{C-to-HLS-C conversion} (107 each)} \\
\hspace{0.5em}C2HLSC, CFull107 rerun~\cite{zhao2026evidencedrivenllmagentctosynthesizablec} & All-pass & 34.6 \\
\hspace{0.5em}Zhao et al., CFull107, gpt-5-mini~\cite{zhao2026evidencedrivenllmagentctosynthesizablec} & All-pass & 87.9 \\
\hspace{0.5em}Zhao et al., CFull107, DeepSeek v4-pro~\cite{zhao2026evidencedrivenllmagentctosynthesizablec} & All-pass & 95.3 \\
\hspace{0.5em}\textsc{FPGA-GEN}, HLS-LeVeri & Admission & 97.2 \\
\bottomrule
\end{tabular}
\end{minipage}
\end{center}
These unmatched endpoints do not attribute gaps to \textsc{FPGA-GEN}.

\textbf{Validity.} Budgets are 10 samples/task (agentic),
five (ours) and 20 (ChatHLS). Models, tasks and endpoints differ; conversion is
cumulative, so this is not a controlled ranking. Tests are not proofs, and
device fit is not routed timing or board execution.

\begin{samepage}
\section{Conclusion}
FPGA-GEN distills FPGA engineering expertise into a bounded workflow for
HLS-C generation. Its contribution is to make requirements explicit across
tasks and functions, and to connect candidate selection and diagnostic repair to tool evidence. Oracle coverage and repeated model and tool execution remain practical
challenges. As model capability and hardware-aware training data improve,
parts of the workflow may become obsolete. Future work should extract
reusable patterns from verified conversion and repair trajectories and test
direct C/C++-to-HLS-C conversion with fewer intermediate stages. Comparisons
should hold generation budgets and validation criteria fixed to distinguish
model improvements from workflow effects. The goal is to transfer more
engineering expertise into the model while preserving functional behavior
and target-specific hardware constraints.

\end{samepage}

\section*{Content Generated by AI}
Codex (GPT-6) and Claude helped review and revise the manuscript. Codex helped
draft the title, abstract, Section~I contributions and Sections~III--V, prepare
selected tables, Fig.~\ref{fig:fpga-gen-workflow} and \LaTeX{}, and analyze
records and check arithmetic. Claude revised the abstract and Section~IV. The
authors cross-checked all AI-assisted content and take responsibility.

\clearpage
\bibliographystyle{IEEEtran}
\bibliography{ieee_bib_control,refs}

\end{document}